\documentclass[manuscript]{aastex631}
\usepackage{bm,amsmath}

\received{September 4, 2024}
\revised{September 12, 2025}
\accepted{September 14, 2025}

\graphicspath{{./}{figures/}}

\begin{document}

\title{Magnitude of Short-Wavelength Electric Field Fluctuations in Simulations of Collisionless Plasma Shocks}

\correspondingauthor{Vadim Roytershteyn}
\email{vroytershteyn@spacescience.org}

\author[0000-0003-1745-7587]{Vadim Roytershteyn}
\affiliation{Space Science Institute, Boulder, CO, USA}

\author[0000-0002-4313-1970]{Lynn B. Wilson III}
\affiliation{NASA Goddard Space Flight Center, 
Heliophysics Science Division, Greenbelt, MD, USA \\
}

\author[0000-0002-4768-189X]{Li-Jen Chen}
\affiliation{NASA Goddard Space Flight Center, 
Heliophysics Science Division, Greenbelt, MD, USA \\
}

\author[0000-0003-1236-4787]{Michael Gedalin}\affiliation{Department of Physics, Ben Gurion University of the Negev, Beer-Sheva, Israel}

\author[0000-0002-6409-2392]{Nikolai Pogorelov}
\affiliation{Department of Space Science, The University of Alabama in Huntsville, AL, USA}
\affiliation{Center for Space Plasma and Aeronomic Research, The University of Alabama in Huntsville, AL, USA}
 
\begin{abstract}
Large-amplitude electrostatic fluctuations are routinely observed by spacecraft upon traversal of collisionless shocks in the heliosphere. Kinetic simulations of shocks have struggled to reproduce the amplitude of such fluctuations, complicating efforts to understand their influence on energy dissipation and shock structure. In this paper, 1D particle-in-cell simulations with realistic proton-to-electron mass ratio are used to show that in cases with upstream electron temperature $T_e$ exceeding the ion temperature $T_i$, the magnitude of the fluctuations increases with the electron plasma-to-cyclotron frequency ratio $\omega_{pe}/\Omega_{ce}$, reaching realistic values at $\omega_{pe}/\Omega_{ce} \gtrsim 30$. The large-amplitude fluctuations in the simulations are shown to be associated with electrostatic solitary structures, such as ion phase-space holes. In the cases where upstream temperature ratio is reversed, the magnitude of the fluctuations remains small.
\end{abstract}


\section{Introduction} \label{sec:intro}

Shocks are ubiquitous in collisionless plasmas and are of great interest due to their role as some of the most efficient energy converters in the universe. Unlike their counterparts in collision-dominated gasses, shocks in collisionless plasmas may dissipate the energy of the upstream flow by a variety of mechanisms as reviewed e.g., by~\citet[][]{sagdeev_1966,biskamp_1973,papadopoulos_1985,Balogh2013,Wilson2014a}. Identification of the specific mechanisms operating under given conditions has been the subject of intense debate and many basic questions remain unresolved. One of such questions is the role of micro-instabilities and large-amplitude fluctuations of the electric field. Spacecraft observations demonstrate that, in contrast to simple theoretical constructs, the shock transition is characterized by the presence of intense fluctuations with magnitude comparable or even significantly exceeding corresponding mean values. Specifically, the focus of the present paper is on intense, short-wavelength fluctuations of the electric field. To a large degree, the present study is motivated by the survey of observational data presented by ~\citet{wilson_2007,Wilson2014a,Wilson2014b}, which indicated that such fluctuations are ubiquitously present, with only weak dependence on the shock parameters. Moreover, estimates of the effective dissipation due to the electric field fluctuations yielded values comparable with the global dissipation rate. Such observations reinvigorated the debate on the role of the fluctuations. 

The difficulties of assessing the contribution of fluctuations to the global dissipation using only limited information obtained along a spacecraft trajectory are self-evident. While the simulations do not suffer from similar limitations, they are hampered by the challenges in resolving the full range of relevant spatio-temporal scales and the sensitivity of the physics of interest to such parameters as the assumed proton-to-electron mass ratio or the separation of scales between the Debye scale and other internal plasma scales, such as ion and electron inertial length and gyroradii. In fact, majority of the self-consistent simulations performed to date have struggled to reproduce observed magnitude of the electric field fluctuations~\citep[e.g.,][]{riquelme_2011, Umeda2012a}, see also a review of the problem by~\citet[][]{Wilson2021} and references therein. Furthermore, parametric studies found strong dependence of the relevant instabilities on numerical parameters of the simulations, such as proton-to-electron mass ratio~\cite[e.g.,][]{Umeda2012b}. In order to make progress, many prior simulation studies have focused on analysis of beam-driven instabilities in local configurations modeling shock foot~\citep[e.g.,][]{shimada_2004,Matsukiyo2006,Muschietti2013,Muschietti2017}. While much has been learned from such studies, they lack self-consistent feedback between macroscopic shock structure and the microscopic fluctuations. 

In this paper, we present results of a scaling study that systematically investigates the dependence of the magnitude of electric field fluctuations in particle-in-cell (PIC) simulations of collisionless shocks on the ratio between plasma frequency $\omega_{pe}$ and the electron cyclotron frequency $\Omega_{ce}$. The simulations are one-dimensional (1D), which suppresses instabilities with a finite component of wavevectors perpendicular to the shock normal. Nevertheless, several key observations could be made: 1) formation of electrostatic solitary structures, {associated with} ion phase-space holes, plays a major role in simulations with the upstream electron temperature $T_e$ exceeding ion $T_i$, such that $T_e / T_i \approx 3$. The magnitude of the electric field fluctuations in such cases is comparable to the observations when normalized to upstream convective electric field, provided the ratio $\omega_{pe}/\Omega_{ce}$ is sufficiently high; 2) the normalized magnitude of the fluctuations is shown to increase with the value of $\omega_{pe}/\Omega_{ce}$; 3) the amplitude of the electric field fluctuations in the case of $T_i/T_e \approx 3$ is significantly smaller than in the case  $T_i/T_e \approx 1/3$.

\section{Numerical method} \label{sec:pic}
We consider 1D simulations of a quasi-perpendicular collisionless shock characterized by upstream Alfv\'en Mach number $M_A = V_0/V_A \approx 10.5$ and the angle between shock normal and the magnetic field $\theta_{Bn}=65^\circ$. Here $V_0$ is the upstream plasma speed and $V_A = B_0/\sqrt{4 \pi n_0 m_i}$ is the Alfv\'en speed defined with the upstream magnetic field $B_0$, density $n_0$, and ion mass is $m_i$. The upstream plasma consists of two populations (ions and electrons) with isotropic Maxwellian distributions characterized by density $n_0$ and temperatures $T_i$ and $T_e$. Unless otherwise specified, the temperatures are chosen such that $\beta_e = 1$ and $\beta_i=0.3$, where $\beta_s = 8\pi n_0 T_s/B_0^2$ and $s=e,i$ denotes particle species. This is consistent with statistical studies of the solar wind at the Earth's heliospheric distances~\citep{wilson_erratum_2023}. The simulations  differ in the value of the ratio between electron plasma and cyclotron frequencies, which is chosen to be  $\omega_{pe}/\Omega_{ce}=\{8,16,32\}$.  All other physical parameters are identical. The simulations are performed in a domain of size $L_x = 30\, d_p$, resolved by a uniform Cartesian grid with the cells size approximately $0.5\lambda_e$, where $\lambda_e$ is the upstream Debye length and $d_p$ is the upstream ion (proton) inertial length. Here $d_s = c/\omega_{ps}$, where $\omega_{ps}^2 = 4 \pi n_0 q_s^2/m_s$, $c$ is the speed of light, and $q_s$ is the charge. The thermal speed is defined as $v_{ts} = \sqrt{2 T_s/m_s}$ and particle gyroradius as $\rho_s = v_{ts}/\Omega_{cs}$. The {parameters of the simulations} are summarized in Table~\ref{tbl:params}. The upstream plasma is injected with the speed {$V_{in} = 7.5 V_A$} from the boundary at $x=L_x$ and the shock is created by interaction of the incoming plasma with the plasma reflected from $x=0$ boundary, which uses a reflecting, perfectly conducting  boundary condition. The shock speed {$V_{sh}$} in the simulation frame is approximately $3 V_A$. The ion-to-electron mass ratio is realistic $m_i/m_e = 1836$. The number of particles per cell $N_\mathrm{ppc}$ corresponding to the upstream density is $N_\mathrm{ppc}=1000$ for each species. No particle splitting/merging algorithms are used in the simulations, so that all of the particles have equal and fixed statistical weight (i.e., charge and mass). Additional simulations with $N_\mathrm{ppc}=4000$ were performed for the case with $\omega_{pe}/\Omega_{ce}=32$. The simulations were performed using the VPIC code~\citep{Bowers2008}. 

\begin{table}[htp]
\caption{Simulation parameters. Here $N_x$ is the number of cells, $\delta x$ is the cell size, and $\delta t$ is the time step.}
\begin{center}
\begin{tabular}{|c|c|c|c|c|c|c|}
\hline
$\omega_{pe}/\Omega_{ce}$  &   $\beta_e$  & $\beta_i$ &    $\delta t \omega_{pe} $ & $N_x$   & $\delta x / d_e$  & $\delta x / \rho_e$  \\
\hline
8                                              &          1    &  0.3         &  0.034       & 28672   & 0.045                 & 0.045 \\
16                                            &           1   &   0.3        & 0.017          & 57344    & 0.022                 & 0.022 \\
32                                            &            1   &    0.3      & 0.008       & 114688  & 0.011                 & 0.011\\
32                                            &           0.3  &   1         &   0.004      & 229376  &  $5.6\times 10^{-3} $  & 0.01  \\
\hline
\end{tabular}
\end{center}
\label{tbl:params}
\end{table}%

While we do not perform a direct comparison of the simulations with any given observation, the chosen upstream parameters are characteristic of the Earth's bow shock or the interplanetary shocks. However, the value of $\omega_{pe}/\Omega_{ce}$ differs in the simulations compared to the realistic ones. Thus a scaling assumption must be made to map the simulation quantities to the observed ones. The shock is specified by the choice of upstream values of dimensionless parameters $M_A$, $\beta_e$, $\beta_i$, and $\theta_{Bn}$. When those are matched between the simulations and the physical system of interest, many other dimensionless ratios are matched as well. These include e.g., the ratio between species inertial length and the gyro-radius, the ratio of a typical convective time scale to the cyclotron time, the ratio between Alfv\'en speed and the ion thermal speed, or the ratio between proton and electron inertial lengths (assuming the chosen mass ratio is realistic, as is the case for the simulations discussed here). Under the chosen scaling, the ratio $\omega_{pe}/\Omega_{ce}$ controls the separation of scales between Debye length and the inertial length/gyroradius and between plasma and the cyclotron frequencies. It is also important to note that both the proton-to-electron mass ratio and the ratio $\omega_{pe}/\Omega_{ce}$ affect the dispersion of whistler precursors~\citep[e.g.,][]{krasnoselskikh_2002}.

\section{Scaling of the electric field fluctuations} \label{sec:scaling}
\begin{figure}[t]
\begin{center}
\includegraphics[width=\textwidth]{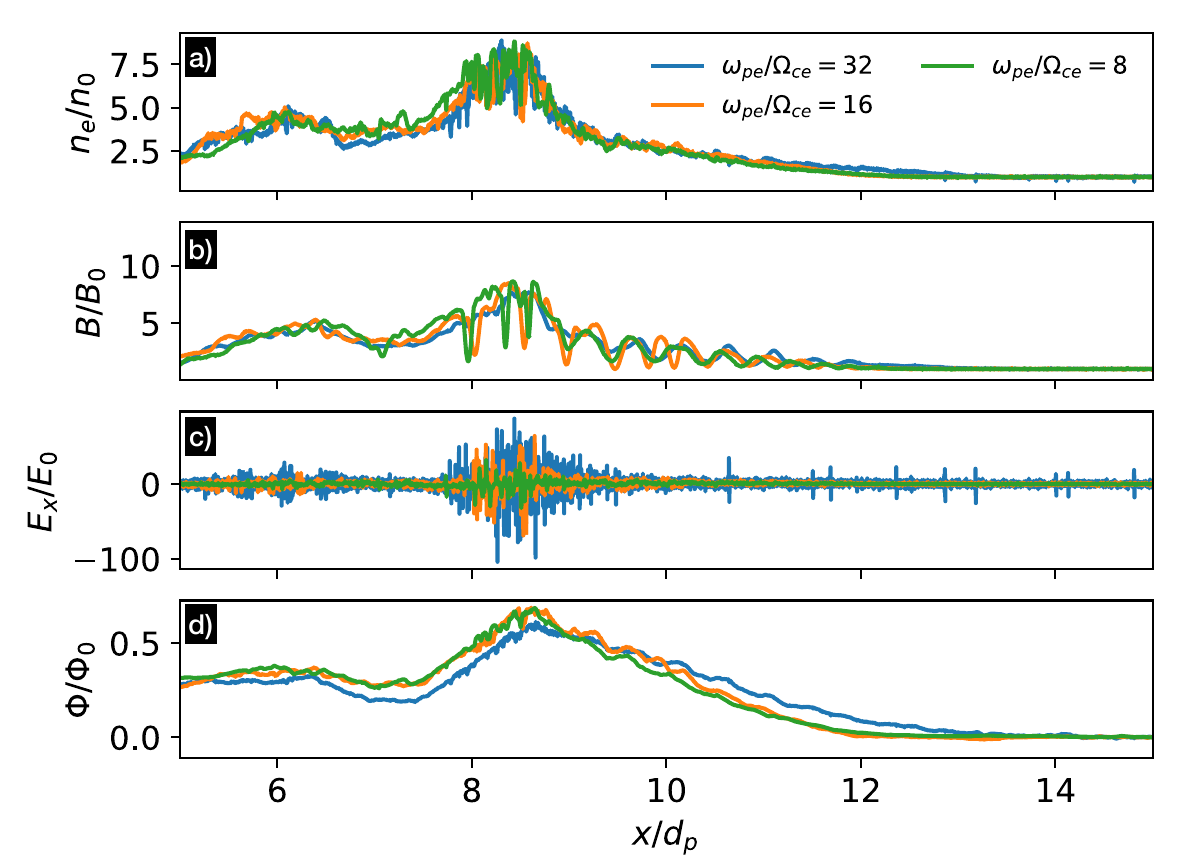}
\caption{Comparison of the simulations with $T_e/T_i\approx 3$ and varying values of $\omega_{pe}/\Omega_{ce}$. Panels a) to d) show profiles of the density, magnetic field, normal component of the electric field, and the NIF shock potential respectively for the cases with $\omega_{pe}/\Omega_{ce}=8,16,32$ at $t\Omega_{ci} \approx 3.5$.}
\label{fig:e_scaling}
\end{center}
\end{figure}

Fig.~\ref{fig:e_scaling} presents an overview of the simulations with $T_e/T_i\approx 3$ and the three different values of $\omega_{pe}/\Omega_{ce}$. Panels a) through d) show profiles of the electron density $n_e$, magnetic field amplitude $B/B_0$, the $x$ component of the electric field $E_x$, and the shock potential respectively, collected at the same instance of time $t\Omega_{ci} \approx 3.5$. The macroscopic profiles of the density and magnetic field are similar between the three simulations. It is worth pointing out that the shocks are non-stationary (reforming). Thus matching of the density and magnetic field profiles implies not only that the shock speed in the simulation frame of reference is close between the three cases, but also that reformation cycles are quite similar. Nevertheless, some differences are clearly present in the structure of the precursors and in the small-scale structure of density fluctuations. 

\begin{figure}[t]
\begin{center}
\includegraphics[width=\textwidth]{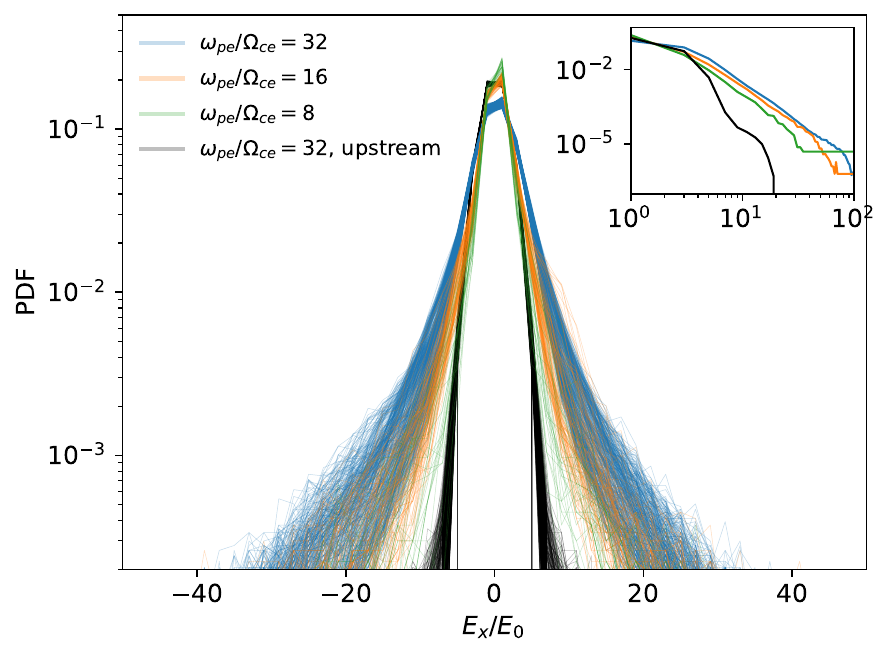}
\caption{Histograms of the electric field fluctuations  $E_x$  in the simulations with $T_e/T_i\approx 3$ (see Fig.~\ref{fig:e_scaling}) during the time interval $t\Omega_{ci} = (2-3)$ in the range $x'/d_p = (3 - 10)$, {where $x'$ is the coordinate co-moving with the shock $x' = x - (t-t_0)V_{sh}$ and $t_0=2\Omega_{ci}$.  The insert shows time-averaged histograms for each case on the log-log scale.}}
\label{fig:e_histogram}
\end{center}
\end{figure}

Panel c) of Fig.~\ref{fig:e_scaling} illustrates one of the principal results of the present study, namely that the magnitude of the electric field fluctuations normalized to the upstream convective electric field $E_0=B_0 (V_0/c) \sin \theta_{Bn} $ increases with the value of $\omega_{pe}/\Omega_{ce}$. The maximum values $E_x \sim 100 E_0$ are achieved in the simulation with  $\omega_{pe}/\Omega_{ce}=32$ (see also Fig.~\ref{fig:e_histogram} discussed below). 

Fig.~\ref{fig:e_histogram} shows  the histogram of normalized electric field $E_x$ computed in the region $x'/d_p=3-10$ for time interval $t\Omega_{ci}=2-3$, which is comparable to the duration of a shock reformation cycle. {Here $x'=x-V_{sh}(t-t_0)$ is coordinate in the frame of reference co-moving with the shock.} Existence of a persistent scaling of $E_x$ with $\omega_{pe}/\Omega_{ce}$ is apparent. Further, there exist a range of values of $E$ where the probability distribution function (PDF) of $E_x$ is characterized by power-law tails such that $\mathrm{PDF} \sim E_x^{-\alpha}$. For example, in the case with $\omega_{pe}/\Omega_{ce}=32$, the power-law exponent of the time-averaged distribution is $\alpha \approx 3.3$ in the range $10 < E_x < 35 $. In the region $E_x \gtrsim 40$, the PDF is much more shallow, with $\alpha \sim 0.4$. However, the statistics is insufficient to reliably establish either the power-law behavior or the extent of this second region at very large values of $E_x$.

\begin{figure}[t]
\begin{center}
\includegraphics[width=\textwidth]{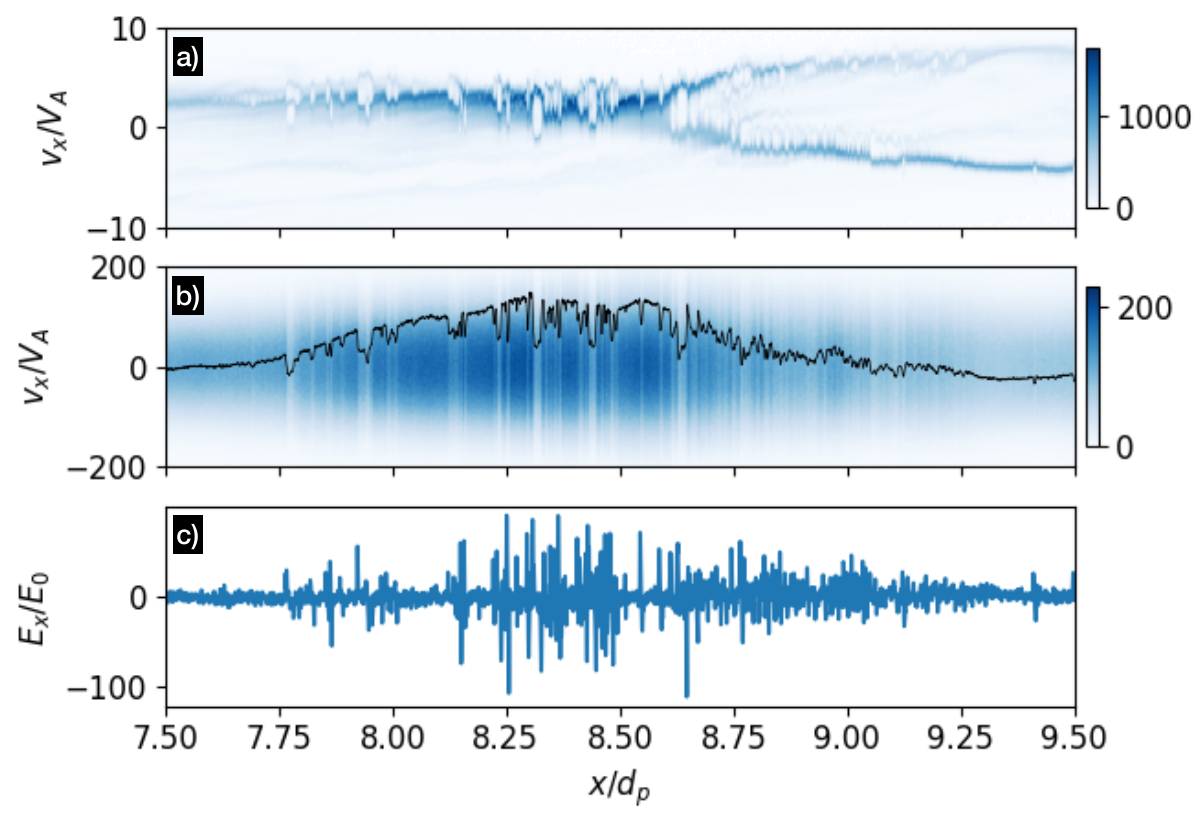}
\caption{ Ion (panel a) and electron (panel b) phase-space structure, together with the associated profile of the electric field (panel c). The black line in panel b) shows variation of the (scaled) electron density across the transition region. }
\label{fig:trapping}
\end{center}
\end{figure}

\begin{figure}[t]
\begin{center}
\includegraphics[width=\textwidth]{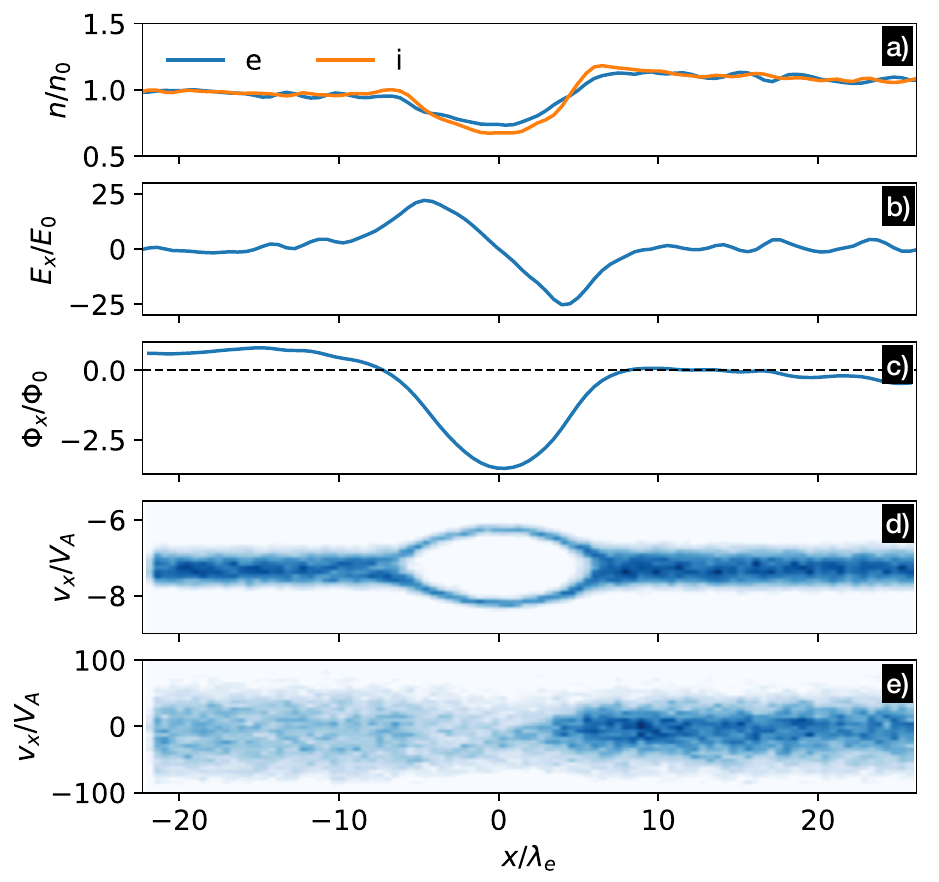}
\caption{An example of ion phase-space hole in the foot region of the shock: profiles of electron and ion densities (a), electric field (b), potential (c) {normalized to $\Phi_0 = T_i/e$}, as well as ion (d) and electron (e) phase space. The spatial coordinates are normalized to the reference value of the Debye length $\lambda_e$ and shifted such that $E_x  \approx 0$ at $x=0$.  }
\label{fig:hole_profile}
\end{center}
\end{figure}

Closer inspection reveals that the largest-amplitude fluctuations of $E_x$ correspond to isolated structures with typical spatial scale of the order of several Debye lengths. {Most of the structures are bipolar and} are associated with ion phase-space holes. This is illustrated in Fig.~\ref{fig:trapping} and Fig.~\ref{fig:hole_profile}. Fig.~\ref{fig:trapping} shows an example of the $(x-v_x)$ phase-space for both electrons and ions in panels a) and b) respectively, together with the profile of $E_x$. The region shown in the figure corresponds to an immediate vicinity of the shock ramp in the simulation with $\omega_{pe}/\Omega_{ce}=32$ at time {$t\Omega_{ci}\approx3.5$}. Characteristic signatures of ion holes are clearly visible in the incoming, reflected, and downstream ion populations. In contrast, no clear phase-space signatures of these structures are seen in the electron distributions, except for variations in total density. In addition to ion phase-space holes associated with individual bipolar $E_x$ structures, some of the ion holes  with the largest spatial extent appear to correspond to ion deflection and possibly trapping  between the structures. 

The ion phase-space holes can also form in the foot region of the shock. They are detectable as large-amplitude bipolar structures in $E_x$, as seen for example in Fig.~\ref{fig:e_scaling}, panel c) around $x\sim 13.2 d_p$. An example of an ion hole observed in the foot region is shown in Fig.~\ref{fig:hole_profile}. The structure is characterized by an overall density depletion and is negatively charged, as can be seen in panel a).  The depth of the potential well is approximately {$3.4 T_i \sim T_e$}. The distance between the two peaks of the $E_x$ profile, shown in panel b), is approximately $10 \lambda_e \sim 0.2 d_e$. Panels d) and e) illustrate the ion and electron phase-space respectively, The structure moves with respect to local plasma flow with a speed of approximately $0.2V_A$, substantially slower than the sound speed $C_s = \sqrt{T_e/m_i} \approx 0.7 V_A$. 

\begin{figure}[t]
\begin{center}
\includegraphics[width=\textwidth]{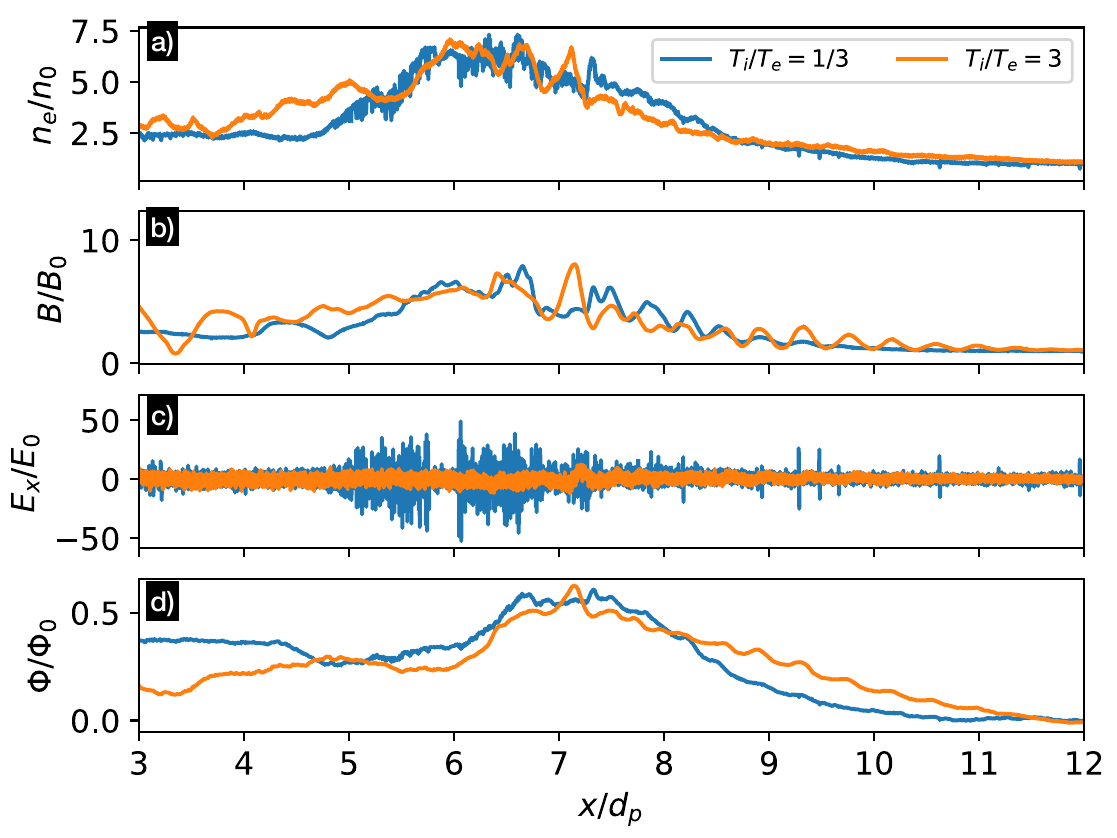}
\caption{Comparison between two otherwise identical simulations with $T_e/T_i \approx 3$ and $T_e/Ti \approx 1/3$. Figure format is the same as in Panels a)-d) of Fig.~\ref{fig:e_scaling}. Slightly different times were used for comparison, since the shock speed differs between the two cases.}
\label{fig:t_dependence}
\end{center}
\end{figure}

In contrast to the cases discussed above, simulations with reversed ratio of electron-to-ion upstream temperature do  not show fluctuations of the electric field of comparable amplitude. This is illustrated in Fig.~\ref{fig:t_dependence}, which compares two otherwise identical simulations i) $\beta_e = 0.3$ and $\beta_i=1$  and ii) $\beta_e = 1$ and $\beta_i=0.3$ in a format similar to Fig.~\ref{fig:e_scaling}. In both cases  $\omega_{pe}/\Omega_{ce}=32$. It is clear that only the simulation with $T_e/T_i > 1$ exhibits large fluctuations of the electric field. Further, the PDF of $E_x$ (not shown) does not develop power-law tails in this case. We also note that the difference in the shock structure between the  cases with $\omega_{pe}/\Omega_{ce}$ fixed and varying {$T_e/T_i$} is more pronounced, compared to the cases with fixed $T_e/T_i$, but varying $\omega_{pe}/\Omega_{ce}$. This is seen for example in density and magnetic field profiles shown in panels a) and b) of Fig.~\ref{fig:t_dependence} respectively.  The time evolution of the shock, including shock speed in the simulation frame, {also differs slightly} between the two cases.

\section{Contribution of the fluctuations to the energy exchange}

\begin{figure}[t]
\begin{center}
\includegraphics[width=\textwidth]{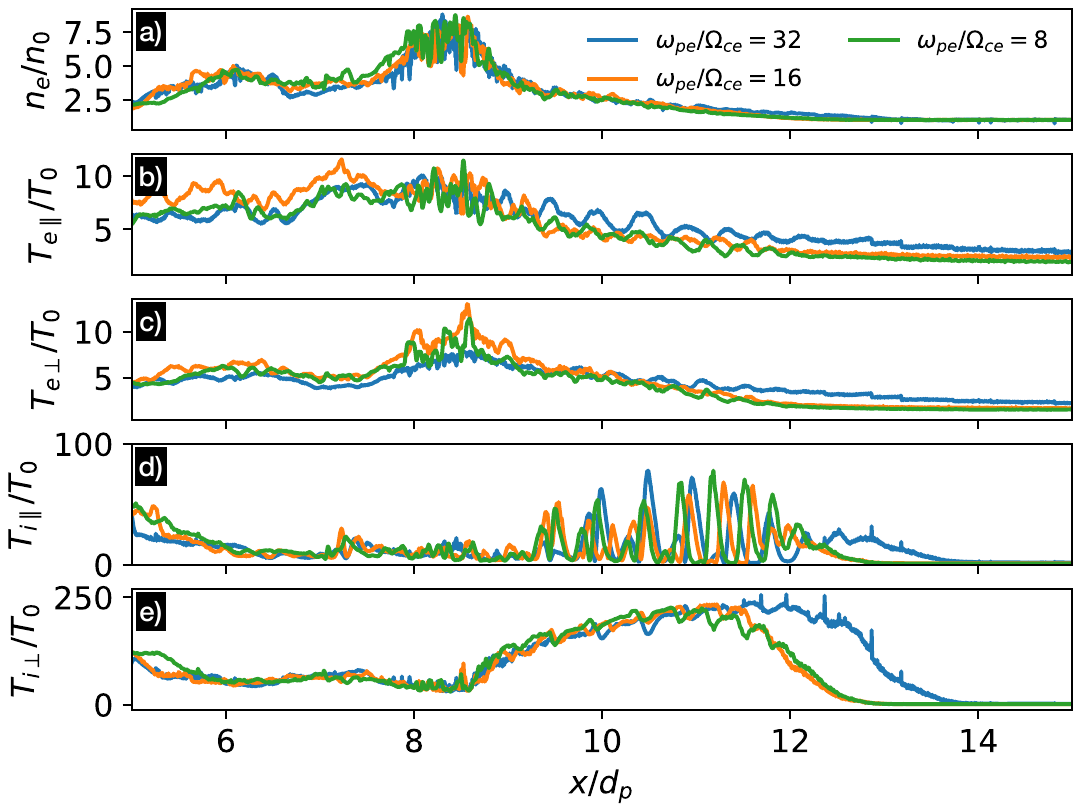}
\caption{Variations of the electron density (a), parallel (b),(d) and perpendicular (c),(e)  electron kinetic temperatures for electrons and ions respectively for the same simulations and time instance as in~Fig.~\ref{fig:e_scaling}. }
\label{fig:energies}
\end{center}
\end{figure}

While 1D simulations discussed here clearly do not allow the full range of instabilities and the associated electric field fluctuations to develop, it is of interest to investigate whether the large-amplitude fluctuations that are present impact average energetics of the shock. A certain insight can be obtained by comparing the evolution of average particle energies through the shock. This is presented in Figure~\ref{fig:energies}, which shows profiles of the parallel and perpendicular electron and ion kinetic temperatures, which are defined via the second moments of the entire particle distributions in the frame of reference co-moving with the plasma (i.e., the incoming populations are not separated from the reflected ones). The parallel and perpendicular directions are defined with respect to the local magnetic field. We observe that the downstream ion temperatures are similar for all simulations. More variation with $\omega_{pe}/\Omega_{ce}$ is observed in the electron temperature, especially in the perpendicular component.  Further, significant local jumps in $T_i$ and especially in $T_e$, which are associated with the solitary structures, could be observed in the foot region of the shock. Substantial differences in both ion and electron energies between the simulation with $\omega_{pe}/\Omega_{ce}=32$ and the other cases can be seen in the foot region. However, those are likely related to the differences in overall structure of the shock at a given instance in time, rather than to the effect of small-scale electrostatic fluctuations. 

A more discerning analysis of the role played by the fluctuations can be  performed by measuring the magnitude of the individual terms in the energy and momentum conservation laws associated with the shock transition. These relations are summarized briefly in the Appendix. We note that while the shocks discussed in this paper are manifestly non-stationary, the conservation laws for energy and momentum fluxes should still hold in the average sense, after averaging over sufficiently long time intervals. For example, conservation of the energy flux for plasma species $s$ reads
\begin{equation}
   \sum_s \nabla \cdot \left \langle  \bm \chi_s + \bm U_s \cdot \bm p_s  + \bm U_s \varepsilon_s + \bm U_s \frac{m_s n_s U_s^2}{2} \right \rangle = \langle \bm E \cdot \bm j \rangle
   \label{eq:exch1}
\end{equation}
where $\bm \chi_s$ is the heat flux, $\bm U_s$ is the plasma flow, $\bm p_s$ is the pressure tensor, $\varepsilon_s =  \mathrm{Tr}\, \bm p_s/2$, $\bm j$ is the current, and $\langle \ldots \rangle$ denotes time-average (see Appendix for more details and definitions). The term $\langle \bm E \cdot \bm j \rangle $ describes energy exchange with the electromagnetic field and can be expressed in terms of the average Poynting flux $\bm S$ through $\langle \nabla \cdot \bm S \rangle = - \langle \bm E \cdot \bm j \rangle $. It is also useful to consider the form of conservation laws that partially separates fluxes of flow and thermal energy of individual species $s$ (as discussed in more details in the Appendix):
\begin{equation}
\nabla \cdot \left \langle \bm p_s \cdot \bm U_s + \bm U_s \frac{m_s n_s U_s^2}{2}  \right \rangle = \left \langle q_s n_s \bm U_s \cdot \bm E\right \rangle   + \left \langle \bm p_s : \nabla \bm U_s   \right \rangle 
   \label{eq:exch2}
\end{equation}
and
\begin{equation}
\nabla \cdot \left \langle \bm \chi_s  + \bm U_s \varepsilon \right \rangle = -\left \langle \bm p_s : \nabla \bm U_s \right \rangle .
   \label{eq:exch3}
\end{equation}
While interpretation of the energy exchange and dissipation in collisionless plasma is notoriously ambiguous, these relations highlight the role of the pressure-strain interaction term $\left \langle \bm p_s : \nabla \bm U_s \right \rangle$  in mediating conversion of flow energy into thermal energy of plasma species and that of the $\left \langle \bm E\cdot \bm j \right \rangle$ term in mediating energy exchange between flow energy of plasma species and the electric field~\citep[e.g.,][]{Yang2017}. 

\begin{figure}[t]
\begin{center}
\includegraphics[width=\textwidth]{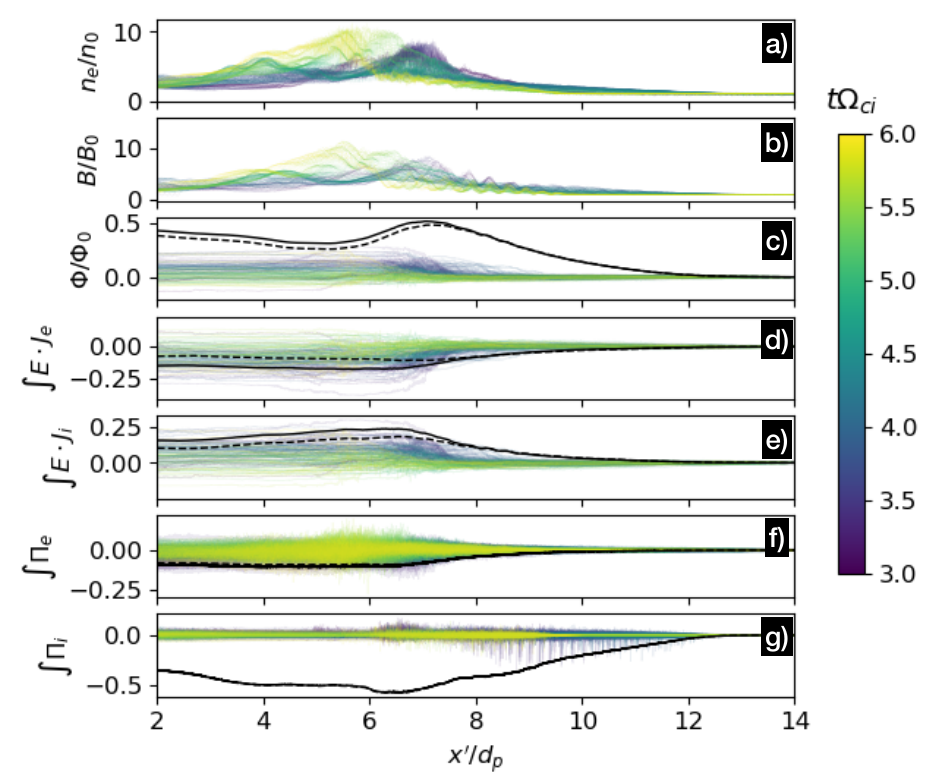}
\caption{Energy exchange terms in the simulation with $T_e/T_i=3$ and $\omega_{pe}/\Omega_{ce}=32$: a) electron density; b) magnetic field; c) NIF shock potential; d) and e)  energy exchange term $\mathcal R = \int^x_{x_0} (\bm  E\cdot \bm j) \, d x$ for electrons and ions respectively; f) and g) the term $\Pi = \int^x_{x_0} (\bm p \cdot \nabla \bm U) \, d x$ for for electrons and ions respectively. For each quantity $A=\{\mathcal R_{e}, \mathcal R_i, \Pi_e, \Pi_i\}$ in panels c) through g), the thin color lines show the difference $\delta A = A - \hat A$, where $\hat A$ is computed with spatially smoothed quantities, the solid black lines show time-average of $A$, and the dashed lines the time-average of $\hat A$. Color corresponds to the simulation time as indicated by the color bar. See text for more details.}
\label{fig:edotj}
\end{center}
\end{figure}

We note that neither the spatial extent, nor the duration of the simulations discussed here are sufficient to establish the values of time-stationary fluxes with sufficient accuracy to carefully examine the conservation laws. Nevertheless, it is instructive to examine relative magnitude of various energy exchange terms in Eq.~\ref{eq:exch1},\ref{eq:exch2},\ref{eq:exch3} and their variation with time. These are presented in Fig.~\ref{fig:edotj}. For reference, panels a) and b) show evolution of the electron density and magnetic field during time interval $t\Omega_{ci} = (3-6)$, which corresponds approximately to a reformation cycle. In these and subsequent panels, individual profiles have been shifted into the normal incidence frame (NIF) using, in addition to Lorentz transformations, a coordinate transformation $x' = x - V_{sh} (t - t_0)$, where {$V_{sh}=3 V_A$} is the approximate average shock speed in the simulation frame and $t_0$ is the beginning of the time interval. In order to assess the influence of the fluctuations on shock properties, we note that most of the intense fluctuations are highly localized and therefore could be removed by spatial averaging. We emphasize that the following analysis relies on this important assumption, which in effect equates an average over short  spatial scales to an average over fast temporal scales. Panels c) shows evolution of quantity $\Phi - \hat \Phi$, where $\Phi = -\int^x_{x_0} E_x d x$ is the NIF shock potential and $\hat \Phi$ is the same quantity computed from a spatially-averaged electric field $\bar E_x$. The reference point for the integration, where $\Phi=0$, is chosen upstream, at $x' = 14 d_p$. A median filter of width $1d_e$ is used to perform the spatial averages. The solid black lines in panel c) of Eq.~\ref{eq:exch1} shows the time-average of the shock potential, while the dashed line shows the time-average of the smooth potential $\hat \Phi$. We observe that, instantaneously, fluctuations can make a substantial contribution to the shock potential. The differences between time-averaged potential and its smooth counterpart are however substantially smaller and appear to originate mostly from the immediate vicinity of the shock ramp. It is worth noting in that regard that the spatial average also smooths of the  ramp.

Using a similar presentation format, panels d) and e) of Fig.~\ref{fig:edotj} examine contribution of the fluctuations to the energy exchange terms for electrons and ions respectively. Specifically, panel d) shows instantaneous difference $\delta \mathcal R_e = \mathcal R_e - \hat {\mathcal R}_e $ between $\mathcal R_e = \int^x_{x_0} \bm E \cdot \bm j_e \, dx$ and $\hat {\mathcal R}_e = \int^x_{x_0} \bar {\bm E} \cdot \bar {\bm j_e} \, dx$ (thin lines). The color of individual lines in panel d) corresponds to the simulation time, as indicated by the color bar. Both quantities are normalized to the upstream energy flux $m_i U_0^3/2$. The solid and dashed black lines show time-averages of {$\mathcal R_e$ and $\hat {\mathcal R}_e$} respectively. The quantity $\delta \mathcal R$ can be interpreted as an estimate of the contribution of short-wavelength fluctuations to the energy exchange term $\mathcal R$. We  observe that instantaneous values of $\delta R_{e,i}$ can be as large as 20\% of the incoming energy flux and can be as large as, or even exceed, the corresponding time-averaged values of $\mathcal R_{e,i}$. The time-averaged contribution of the fluctuations to the energy exchange is smaller, of the order of 5\% of the incoming energy flux.  

Similar analysis for the quantity $\Pi_s = \int^x_{x_0} (\bm p : \nabla \bm U) d x$ is shown in panels f) and g) for electrons and ions respectively. It is clear that the fluctuations can significantly affect or even dominate the instantaneous values of $\Pi_s$, as seen in the foot region of the shock. Their time-average contribution however is much smaller. 

\section{Discussion}
The simulations discussed in this paper demonstrate several noteworthy results. First, they explicitly show that the amplitude $E$ of the electrostatic fluctuations generated in the vicinity of the shock transition region can reach high amplitudes, such that $E/E_0 \sim 100$. Such values are achieved provided that the proton-to-electron mass ratio and the the value of $\omega_{pe}/\Omega_{ce}$ are simultaneously large. For the shock parameters  considered in the present study, $m_i/m_e=1836$ and $\omega_{pe}/\omega_{ce} \gtrsim 30$ were required.  Here $E$ is the characteristic amplitude of the fluctuations and $E_0$ is the upstream convective electric field defined above. Such large-amplitude fluctuations are routinely observed in the spacecraft data, but have not been previously reproduced by self-consistent shock simulations. Thus the presented results resolve (with some caveats related to dimensionality of the simulations, see below) the inconsistency between simulations and observations discussed in a brief review by~\citet{Wilson2021}. The existence of such a scaling could be easily surmised under some reasonable assumptions on the saturated amplitude of the potential and characteristic scale length of the fluctuations (see below). However, it is only one of several possibilities and, to the best of our knowledge,  the scaling had not been explicitly demonstrated before. Furthermore, the simulations suggest existence of a strong dependence of the magnitude of the electric field fluctuations on the ratio between upstream electron and ion temperature. 

Second important observation is that in the present simulations the large-amplitude electric field fluctuations correspond to electrostatic solitary structures. {Most of the structures are bipolar} with negative potential (ion phase-space holes), {although some unipolar spikes in the electric field are also present.} The structures form throughout the shock transition region {and correspond to a complex, multi-scale pattern in the ion phase-space.} In the foot of the shock, they are  capable of trapping ions and are seen to move with respect to the upstream flow with velocities that are a fraction of the sound speed. {Existence of such fluctuations is not in itself a new result. They have been reported in the very early shock simulations~\citep[e.g,][]{Biskamp_1972}. The novel result is rather their persistence at nearly realistic simulation parameters, their ability to produce fluctuations of electric field with amplitudes comparable to the observed, and the analysis of their influence on electron and ion energetics. Further,} ion holes have been identified as the dominant (i.e., the likeliest to be detected) electrostatic solitary structures associated with the Earth's bow shock~\citep[][]{Vasko2018,Wang2020,Wang2021}. They are typically assumed to originate from streaming instabilities~\citep[e.g.,][]{shimada_2004}. {Unipolar structures (double layers) have also been detected~\citep[e.g.,][]{Sun_2022}.} The abundance and stability of the {structures} are almost certainly influenced by the dimensionality of the simulations. In higher dimensions, other types of instabilities {such as ion sounds waves, which are often detected in observations \citep[e.g.,][]{Vasko_2022}}, will  develop and will likely compete with formation of {the solitary structures}. Further, the {solitary structures} may become unstable. We anticipate however that the basic result, namely increase in the characteristic amplitude of the fluctuation with $\omega_{pe}/\Omega_{ce}$ will remain valid.  

{We note that we have not been able to clearly identify the generation mechanism of the solitary structures observed in the simulations. An ion-ion instability associated with the reflected ion beam is the likely mechanism in the foot region of the shock. However, the structures are detected across a wide transition region, including far in the foot and downstream of the ramp. There is some circumstantial evidence that the generation mechanism is nonlinear, at least in the upstream regions. We defer a detailed investigation of this issue to future investigations.}

The observed scaling of $E$ could be understood under an assumption that the spatial extent $\ell$ of typical electrostatic structures is of the order of the Debye length. Indeed, the magnitude of the potential differential {$\Delta \phi$} associated with structures must be related to a characteristic energy scale, such as electron $T_e$ or ion $T_i$ temperatures or the upstream ion energy $m_i V_0^2/2$. The specific scaling of {$\Delta \phi$} depends on the mechanisms of the hole formation and/or its stability, neither of which are addressed in the present work. Regardless of these details, the resulting scaling is $E/E_0 \propto  \omega_{pe}/\Omega_{ce}$.  For example, assuming $\Delta \phi \sim T_e$ yields $E/E_0 \sim (\Delta \phi/\ell) ( c/V_0 B_0) \sim \sqrt{\beta_e/2} \sqrt{m_i/m_e} (\omega_{pe}/\Omega_{ce}) / M_A$. Similarly, assuming that {$\Delta \phi$} is related to the upstream ion energy (which could be a relevant estimate for the shock ramp) yields $ E/E_0 \sim M_A (\beta_e/2)^{-1/2}  \sqrt{m_i/m_e} (\omega_{pe}/\Omega_{ce}) $. In contrast, assuming $\ell \sim d_e$ or $\ell \sim \rho_e$ yields scaling for $E/E_0$ that is independent on $\omega_{pe}/\Omega_{ce}$. 

We note that in a recent work,  ~\citet{Bohdan2024} suggested that in low-Mach number shocks the amplitude of electrostatic fluctuations associated with electron acoustic waves (EAWs) and related electrostatic solitary structures in the form of electron phase-space holes can reach values comparable to the observed. This assessment was based on extrapolating a scaling of the amplitude of electric field fluctuations with the ratio of the shock speed to the speed of light, which is equivalent to scaling $\omega_{pe}/\Omega_{ce}$ in our discussion.  This result, complimentary to {the findings summarized here}, offers an attractive possibility for explaining ubiquitous presence of high-amplitude electrostatic fluctuations in observations, with only weak dependence on shock parameters such as the shock speed {or even the electron to proton temperature ratio}.

Finally, an assessment of the influence of fluctuations developed in the present simulations on various energy exchange channels was performed, based on the analysis of their contribution to quasi-stationary fluxes of energy. This analysis demonstrated that, instantaneously, the fluctuations may produce a contribution comparable {to} or even exceeding time-averaged fluxes. When time-averaged, the contribution of the fluctuations was substantially smaller. However, it is important to note that i) the accuracy of the analysis was constrained by the spatial extent and time duration of the simulations, such that accurate values of the quasi-stationary fluxes cannot be established;  ii) the 1D nature of the simulations constrains the types of instabilities that can develop, leaving out for example such important modes as the ion acoustic fluctuations; and iii) the maximum value of $\omega_{pe}/\Omega_{ce}=32$ used in the simulations is still substantially smaller than $\omega_{pe}/\Omega_{ce} \sim 100$, which is characteristic of typical heliospheric shocks at 1AU. A full assessment of the influence of microinstabilities on shock structure and energy exchange requires high-resolution, large-scale multidimensional simulations with large values of $\omega_{pe}/\Omega_{ce}$. This is a formidable computational task for existing simulation codes.  Nevertheless, understanding of how the presented results extend to higher-dimensional cases is a crucial issue.

\begin{acknowledgments}
VR was partially supported by NASA grant 80NSSC21K1680 and by NSF grants 2010144 and 2512084. VR and MG were partially supported by the International Space Science Institute (ISSI) in Bern,  through International Team project \#23-575. L.C. was supported by the NASA MMS mission. N.P. was supported, in part,
by NSF-BSF award 2010450. Some of the work was supported by the Geospace Environment Modeling (GEM) Focus Group entitled, "Particle Heating and Thermalization in Collisionless Shocks in the Magnetospheric multiscale mission (MMS) Era," led by L.B. Wilson III.  Computational resources were provided by the the NASA High-End Computing (HEC) Program through the NASA Advanced Supercomputing (NAS) Division at Ames Research Center and by Texas Advanced Computing  Center (TACC) at the University of Texas at Austin. Access to TACC resources was provided by Frontera Pathways allocation PHY23034. Supporting simulations were also performed using resources of the National Energy Research Scientific Computing Center (NERSC), a U.S. Department of Energy Office of Science User Facility located at Lawrence Berkeley National Laboratory, operated under Contract No. DE-AC02-05CH11231 using NERSC award FES-ERCAP0024215. 
\end{acknowledgments}

\vspace{5mm}

\software{VPIC \citep{Bowers2008,VPICgit}
          }

\appendix
\section{Energy and momentum conservation through a collisionless shock}

The exact conservation laws for Vlasov-Maxwell system describing collissionless plasma are readily obtained by taking moments of the Vlasov equation. Specifically, the momentum conservation equation for each plasma species $s$ reads
\begin{equation}
    \partial_t \bm M_s + \nabla \cdot \mathbb P_s = \rho_s \bm E  + \frac{1}{c} \bm j_s \times \bm B
    \label{eq:mom1}
\end{equation}
while the energy conservation is
\begin{equation}
    \partial_t  \mathcal{E}_s  + \nabla \cdot \mathbb T_s = \bm E \cdot \bm j_s.
    \label{eq:en1}
\end{equation}
Here the momentum density is
\begin{equation}
    \bm M_s = m_s \int ( \bm v f_s) \, d^3 v =  m_s  n_s \bm U_s,
\end{equation}
where $f_s$ is the distribution function of particles of species $s$,
\begin{equation}
    n_s =  \int ( f_s) \, d^3 v 
\end{equation}
is the particle density, and
\begin{equation}
    \bm U_s =  \frac{1}{n_s}  \int ( \bm v f_s) \, d^3 v 
\end{equation}
is the mean flow. Charge density of species $s$ with charge $q_s$ is $\rho_s = q_s n_s$ and the current is $\bm j_s = q_s n_s \bm U_s$. The particle momentum flux is
\begin{equation}
    \mathbb P_s =  m_s \int ( \bm v \bm vf_s) \, d^3 v =  \bm p_s + m_s n_s \bm U_s \bm U_s,
\end{equation}
the particle energy density is
\begin{equation}
    \mathcal E_s =  \frac{m_s}{2} \int ( v^2 f_s) \, d^3 v =    \varepsilon_s +  \frac{m_s n_s U_s^2 }{2},
\end{equation}
where the pressure tensor is 
\begin{equation}
\bm p_s=  m_s  \int \left ( \bm w_s \bm w_s f_s \right ) \, d^3 w,
\end{equation}
with $\bm w  = v - \bm U$. The particle energy flux is 
\begin{equation}
    \mathbb T_s =  \frac{m_s}{2} \int ( \bm v v^2 f_s) \, d^3 v =  \bm \chi_s + \bm U_s \cdot \bm p_s  + \bm U_s \varepsilon + \bm U_s \frac{m_s n_s U_s^2}{2},
\end{equation}
where $\bm \chi_s$ is the heat flux
\begin{equation}    
\bm \chi_s=  \frac{m_s}{2} \int ( \bm w w^2 f_s) \, d^3 w  
\end{equation}
and $\varepsilon_s$ is the energy density in the comoving frame
\begin{equation}
\varepsilon_s =  \frac{m_s}{2} \int ( w^2 f_s) \, d^3 w     =  \frac{1}{2} \mathrm{Tr} \, \bm p_s.
\end{equation}
The right-hand side in Eqs.~\ref{eq:mom1} and~\ref{eq:en1} are terms describing exchange of momentum and energy between particles and the electromagnetic field. Specifically the field momentum and energy conservation reads,
\begin{equation}
    \frac{1}{c^2}\partial_t \bm S + \nabla \cdot \mathbb F = -\left (\bm E \rho + \frac{1}{c} \bm j \times \bm B \right )
    \label{eq:mom_field}
\end{equation}
and
\begin{equation}
    \partial_t \left( \frac{E^2+B^2}{2}\right) + \nabla \cdot \bm S = -\bm E\cdot \bm j
    \label{eq:mom_field_b}
\end{equation}
where $\rho = \sum_s \rho_s$ and $\bm j = \sum_s \bm j_s$. Here the Poynting flux is
\begin{equation}
    \bm S = \frac{c}{4 \pi} \bm E \times \bm B,
\end{equation}
and the momentum flux of the electromagnetic field, related to the Maxwell stress tensor, is
\begin{equation}
    \mathbb F = \frac{1}{4 \pi} \left [ \frac{B^2 + E^2}{2} \mathbb I - \bm E \bm E  - \bm B \bm B   \right]
\end{equation}

For a time-stationary (in the average sense) planar shock, the normal components of time-averaged fluxes should be conserved across the shock~\citep[e.g.,][]{gedalin_change_2022}
\begin{align}
    \mathbf {n} \cdot \left \langle \sum_s \mathbb T_s + \bm S \right \rangle  = Const \\
    \mathbf {n} \cdot \left \langle \sum_s \mathbb P_s + \mathbb F \right \rangle = Const,
\end{align}
where $\mathbf {n}$ is the shock normal and $\langle \ldots \rangle $ denotes the average over time and over directions perpendicular to the shock normal. 

The momentum conservation could also be written in the form that largely separates the influence of fast, small-scale electrostatic fluctuations and laminar fields
\begin{equation}
    \nabla \cdot \left \langle \sum_s \mathbb P_s + \frac{1}{2} B^2 \mathbb I - \bm B \bm B  \right \rangle  = \left \langle \rho \bm E + \frac{1}{c} \bm E \times \nabla \times \bm E   \right \rangle.
    \label{eq:mom2}
\end{equation}
The $\bm E \cdot \bm j$ term in the energy conservation equation describes both energy exchange with laminar field (in the form of the average Poynting flux) and with fluctuations
\begin{equation}
    \nabla \cdot \left \langle \sum_s \mathbb T_s \right \rangle = \left \langle \bm E \cdot \bm j \right \rangle 
    \label{eq:en2}.
\end{equation}
A certain insight into the influence of the fluctuations can be obtained by writing $\bm E \cdot \bm j$  as
\begin{equation}
    \left \langle \bm E \cdot \bm j \right \rangle = \langle \bm E \rangle \cdot \langle \bm j \rangle + \langle \delta \bm E \cdot  \delta \bm j \rangle
\end{equation}
where $\delta \bm E$ and $\delta \bm j$ are the the fluctuating components of the electric field and of the current, which are assumed to have vanishing averages. 

We note that using the momentum conservation for individual plasma species, one can also obtain an alternative form of the conservation laws~\citep[see e.g.,][]{Yang2017}
\begin{equation}
\nabla \cdot \left \langle \bm \chi_s  + \bm U_s \varepsilon \right \rangle = - \left \langle \bm p_s : \nabla \bm U_s \right \rangle 
\end{equation}
Subtracting this from the full energy conservation relation results in the relation describing evolution of the flow energy
\begin{equation}
\nabla \cdot \left \langle \bm p_s \cdot \bm U_s + \bm U_s \frac{m_s n_s U_s^2}{2}  \right \rangle = \left \langle q_s n_s \bm U_s \cdot \bm E\right \rangle   + \left \langle \bm p_s : \nabla \bm U_s   \right \rangle 
\end{equation}

This form of the equations highlights the role played by the $\bm E \cdot \bm j$ and the pressure-strain interaction term $\bm p_s : \nabla \bm U_s$ in the energy exchange between thermal and flow energy of particles and the fields.  Finally, we note that analysis of the energy exchange could be performed directly in the phase-space, instead of focusing on moments of the distribution function. As an example, an application of such techniques to analysis of particle energization at perpendicular collisionless shock is discussed by~\citet[][]{juno_2021}.

\bibliography{shocks}{}
\bibliographystyle{aasjournal}

\end{document}